# Non-Trivial Non-Radiating Excitation as a Mechanism of Resonant Transparency in Toroidal Metamaterials

V. A. Fedotov [1, 2, *], A. V. Rogacheva [1, 2], V. Savinov [1, 2], D. P. Tsai [3, 4], N. I. Zheludev [1, 2]

[1] Optoelectronics Research Centre, University of Southampton SO17 1BJ, UK
[2] EPSRC Centre for Photonic Metamaterials, University of Southampton SO17 1BJ, UK
[3] Department of Physics, National Taiwan University, Taipei 10617, Taiwan
[4] Research Center for Applied Sciences, Academia Sinica, Taipei 115, Taiwan

**Abstract:** **We demonstrate theoretically and confirm experimentally a new mechanism of resonant electromagnetic transparency, which yields extremely narrow isolated symmetric Lorentzian lines of full transmission in metamaterials. It exploits the long sought non-trivial non-radiating charge-current excitation based on toroidal dipole moment, predicted to generate waves of gauge-irreducible vector potential in the complete absence of scattered electromagnetic fields.**

Engaging strongly resonant interactions allows dramatic enhancement of functionalities of many electromagnetic devices. The resonances, however, can be dampened by Joule and radiation losses. While in many cases Joule losses may be minimized by the choice of constituting materials, controlling radiation losses is often a bigger problem. Recent solutions include the use of coupled radiant and sub-radiant modes yielding narrow asymmetric Fano resonances in a wide range of systems, from defect states in photonic crystals [1] and optical waveguides with mesoscopic ring resonators [2] to nanoscale plasmonic and metamaterial systems exhibiting interference effects akin to electromagnetically-induced transparency [3]. Here we report a new class of artificial electromagnetic media with ultra-narrow resonances, namely toroidal metamaterials, which exploit interference between collocated and coherently oscillating conventional electrical and special toroidal dipolar modes producing characteristically *symmetric* Lorentzian transparency lines. Moreover, such metamaterials are a long-awaited implementation of the *non-trivial* non-radiating charge-current configurations generating waves of gauge-irreducible vector potential in the absence of electromagnetic fields that were predicted by Afanasiev and Stepanovsky in 1994 [4].

Toroidal dipole is a part of an independent family of multipoles that is complementary to the conventional electric and magnetic multipoles. While electric (charge) and magnetic dipoles can be visualised as a pair of opposite charges and a circular current respectively, toroidal dipole results from currents **j** flowing on the surface of a torus along its meridians (poloidal currents), as illustrated in Fig. 1a. Its moment is given by

$$\mathbf{T} = \frac{1}{10c} \int \left[ (\mathbf{r} \cdot \mathbf{j})\mathbf{r} - 2r^2 \mathbf{j} \right] d^3 r,$$

where **r** is the coordinate vector with its origin placed in the center of the torus [5]. Dynamic excitations of toroidal dipole and higher toroidal multipoles were overlooked for some time as their manifestations in classical electrodynamics are often masked by much stronger electric and magnetic multipoles [6-8]. Only recently, a spectrally isolated transmission resonance corresponding to the toroidal excitation has been observed in an artificially engineered electromagnetic material (metamaterial), where contributions from electric and magnetic multipoles were deliberately suppressed by design [9]. Since then microwave and optical toroidal resonances were identified in a number of metamaterial and plasmonic systems [10-13].

Toroidal and electric dipolar emissions have the same angular momentum and parity properties [8]. Thus, when the poloidal current mode and bipolar charge distribution corresponding to toroidal and electric dipoles respectively are collocated and coherently oscillate at the same frequency, their radiated electromagnetic fields can interfere destructively and disappear completely outside the resulting charge-current configuration, as was first pointed out by Afanasiev and Stepanovsky [4], and demonstrated with further numerical studies [14]. This gives an interesting opportunity of creating artificial electromagnetic materials, which exhibit a very narrow isolated transparency band when the radiation losses are suppressed through the interference of toroidal and electric dipolar emission. The underpinning mechanism of the transparency and resonant dispersion here are somewhat different from those typically encountered in the plasmonic and metamaterial systems mimicking electromagnetically-induced transparency, where the interference between two spectrally separated multipolar modes produces a sharp asymmetric Fano resonance on the background of another much wider resonant band [3].

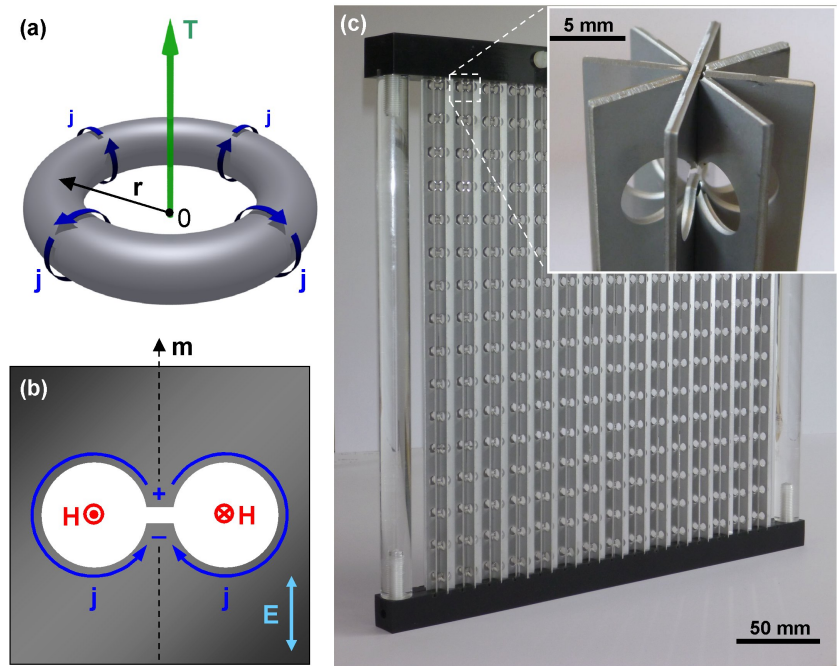

**Figure 1. (Colour online) Toroidal metamaterial.** (**a**) Poloidal currents flowing on a surface of a torus along its meridians create toroidal dipole moment **T**. (**b**) Metal screen with a dumbbell-shaped aperture is the structural element of toroidal metamaterial. Dashed arrow **m** represents axis of its mirror symmetry. (**c**) Photograph of the assembled metamaterial slab, an array of 15 × 16 toroidal aperture-based metamolecules. Inset shows a close-up view of one of the array's column with 8-fold symmetry.

We demonstrate this novel mechanism of radiation suppression for metamaterials based on a dumbbell-shaped aperture element. It is made in a thin metal plate and resembles a meridianal cross-section of a toroidal void (see Fig. 1b). This is a very special electromagnetic system in which the incident wave $\mathbf{E}(t) = \mathbf{E}_0\, e^{i\omega t}$ polarized parallel to the symmetry axis of the aperture **m** induces both electric **P** and toroidal **T** dipolar moments. Indeed, the electric field of the wave drives charge separation $\rho(t) = \rho_0\, e^{i\omega t}$ across the waist of the aperture, which gives rise to an oscillating electric dipole moment $\mathbf{P}(t) \sim \mathbf{m}\rho_0\, e^{i\omega t}$ oriented along **m** (see Fig. 1b). The charge displacement also gives rise to counter-rotating (poloidal-like) currents $j(t)$ oscillating along the edges of the circular cuts, as shown in Fig. 1b. The poloidal currents are proportional to the time derivative of the charge displacement and produce toroidal moment $\mathbf{T}(t) \sim \mathbf{m}\, i\omega\rho_0\, e^{i\omega t}$ that is also oriented along **m** and oscillates coherently with the electric dipole lagging a quarter of the period behind the latter. Now, if the field radiated by the electric dipole $E_P(t)$ is proportional to the first time derivative of $P(t)$, $E_P(t) \sim i\omega\rho_0\, e^{i\omega t}$, the field radiated by the toroidal dipole is proportional to the second time derivative of $T(t)$ [4], i.e. $E_T(t) \sim -i\omega^3 \rho_0 e^{i\omega t}$. Therefore $E_P(t)$ and $E_T(t)$ scattered by the dumbbell-shaped aperture are in anti-phase and always interfere destructively. The amplitude of the poloidal currents increases resonantly when the wavelength of incident radiation $\lambda = 2\pi c/\omega$ becomes close to the circumference of the aperture $\sim 4\pi R$ (where $R$ is the radius of the circular cuts) leading to the enhancement of toroidal emission, which in principle can be made to cancel electric dipolar scattering completely.

In reality, the incident electromagnetic wave induces in such a structure higher oscillating multipoles as well, most notably magnetic quadrupole $\mathbf{Q}_m$, which also results from the pair of counter-rotating currents. The scattering contribution from this multipolar current mode can be effectively suppressed in an aperture-based structure of higher rotational symmetry such as, for example, 4-fold or 8-fold symmetric toroidal metamolecules shown in Figs. 2a and 2b. We demonstrated this by modeling numerically the interaction of the metamolecules assembled in two-dimensional arrays (slabs of toroidal metamaterials) with a normally incident linearly polarized plane wave. The slabs were described through periodic boundary conditions applied to the corresponding unit cell's facets in X and Y directions, as indicated in Figs. 2a and 2b. The dumbbell-shaped aperture resonators forming 4-fold and 8-fold symmetric toroidal metamolecules were assumed to be cut in infinitely thin sheets of perfect electric conductor in accordance with the design specifications. Electromagnetic response of the metamaterial slabs was simulated using commercial full three-dimensional Maxwell equations solver based on the finite element method, COMSOL 3.5a. The simulations also provided data on the densities of electrical currents induced in the metamaterials by the incident wave, which was used to compute scattered powers of the conventional multipoles and toroidal dipole associated with each metamolecule.

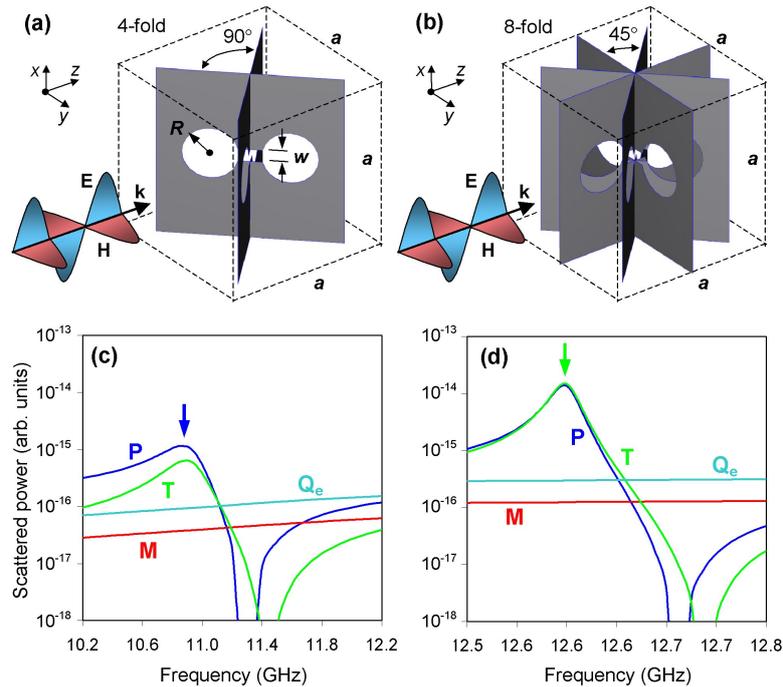

**Figure 2. (Colour online) Multipole excitations in toroidal metamaterials.** Panels **(a)** and **(b)** show metamaterial's unit cell with 4-fold (a) and 8-fold (c) symmetry; $a = 16.5$ mm, $R = 2.5$ mm, $w = 1.2$ mm, separation between the centers of the circular cuts is 7.0 mm. Panels **(c)** and **(d)** show dispersions of multipolar scattering rates calculated for 4 strongest multipoles induced in 4-fold (c) and 8-fold (d) symmetric metamolecules. Arrows indicate locations of the corresponding transparency resonances.

Our calculations showed that, for example, in the metamaterial with 4-fold symmetric metamolecules the emission of the standard multipoles other than electric dipole **P** is small: close to the resonance (which is at around 11 GHz for a centimetre-sized metamolecule) electric quadrupole $\mathbf{Q}_e$ and magnetic dipole **M** exhibit scattering rates that are factors of 10 and 30 smaller than that of **P**, while $\mathbf{Q}_m$ is a factor of $10^5$ weaker here (see Fig. 1c). Toroidal dipole provides the second strongest contribution at the resonance being responsible for more than 30% of the total scattering. Its presence can be detected in the near-field as closed loops of magnetic field-lines confined within the metamolecule and threading the circular sections of the apertures (see left inset to Fig. 3).

Our calculations also show that at the resonance the complete destructive interference of radiated fields take place in a loss-less toroidal metamaterial (metals can be treated loss-less in the microwave part of the spectrum). This corresponds to the total transparency of the toroidal metamaterial. For instance, for the 4-fold system the prevailing electric dipolar scattering is cancelled on about 35% by toroidal scattering and on 65% by scattering from other multiples, creating a background-free resonance transparency peak with $Q = 35$ and Lorenzian profile (see Fig. 3). In the metamaterial composed of 8-fold symmetric metamolecules electric and toroidal dipolar scattering is mutually cancelled on 96%, leading to a transparency peak with $Q \sim 520$. The increased role of toroidal dipolar mode in the 8-fold symmetric structure is also evident from the calculated magnetic near-field map (compare the insets to Fig. 3).

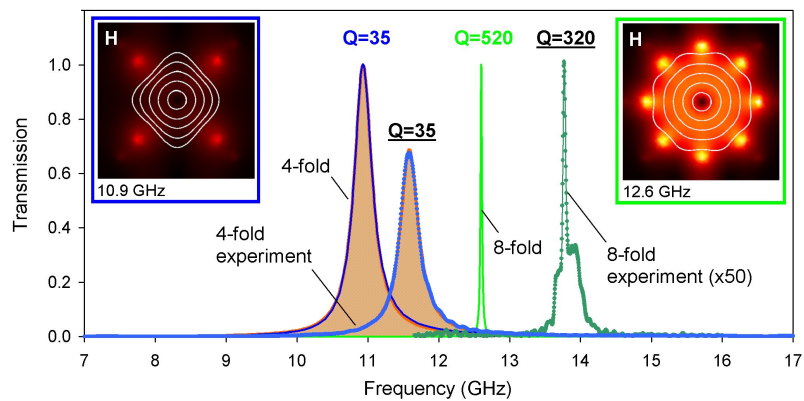

Figure 3. (Colour online) **Transmission response of toroidal metamaterials.** Blue and green solid curves show simulated spectra of metamaterial slabs based on 4-fold and 8-fold symmetric metamolecules respectively, dots - experimentally measured data. Orange-filled areas represent symmetric Lorentz line profiles fitting simulation and experimental data for the 4-fold symmetric structure. Insets show simulated distribution of magnetic field lines and magnetic field intensity |**H**| corresponding to resonantly excited modes in 4-fold (left) and 8-fold (right) symmetric metamolecules.

We confirmed the described transparency resonances due to elecric-toroidal dipolar interference with experiments conducted in the microwave part of the spectrum. For that we constructed toroidal metamaterials from strips of thin stainless steel plates with dumbbell-shaped apertures. The stainless steel strips, which had the thickness of 0.8 mm, were patterned using chemical etching and assembled into columns of 4-fold and 8-fold symmetric structures. The metamaterial slabs were formed by 15 such columns, which contained 16 toroidal metamolecules each (see Fig. 1c). All dimensions of the design features were identical to those used in our modelling (Fig. 2). The transmission spectra of the metamaterial samples were measured in Emerson & Cuming microwave anechoic chamber using vector network analyser (Agilent E8364B) and a pair of broadband linearly polarized horn antennas (Schwarzbeck Mess-Elektronik BBHA 9120D) equipped with dielectric lens collimators. The experimental data fully reproduced the main features of the metamaterials' response predicted by the modelling: the appearance of narrow isolated bands of transparency with very high Q-factors (see Fig. 3). The slight blue shift of experimentally observed resonances results from the non-zero thickness of the metal strips, while somewhat lower Q-factor measured for 8-fold toroidal structure we attribute to the inhomogeneous broadening of the high-fidelity resonance resulting from fabrication tolerances and imperfections of the metamaterial assembly. The latter, in combination with the residual divergence of the incident wavefront, is also responsible for the incomplete transparency at the resonance.

Finally, we point out that simultaneous presence of collocated electric and toroidal dipolar excitations in our metamaterial creates a unique situation, where vector potential **A** generated outside the structure can be non-zero even though the emitted electromagnetic fields are completely cancelled via the destructive interference. Indeed, although the electromagnetic field emission characteristics of oscillating **P** and **T** are identical, the vector-potential fields they produce are quite different [4, 15]. Under the condition of total coherent cancellation of their radiation the resulting vector potential in the far-field is given by

$$\mathbf{A}(\mathbf{r}) = -\frac{\mathbf{r}\,k^2}{r^3}(\mathbf{r}\cdot\mathbf{T})e^{-ikr}.$$

This non-zero vector-potential field cannot be removed by a gauge transformation, rendering the corresponding combination of electric and toroidal dipolar excitations a non-radiating charge-current configuration of the intriguing, *non-trivial* type [4, 15]. Such non-radiating configuration has never been realized before. We argue that the resonant transparency of the demonstrated metamaterials is the first example of the manifestation of the non-trivial non-radiating charge-current excitations. Here, due to the two-dimensional periodicity of the metamaterial arrays, the non-compensated component of the vector potential field is localized in the plain of the metamaterial structure. It should be possible to detect in time-dependent Aharonov-Bohm-type experiments, but the feasibility of such studies requires further investigation.

In conclusion, we have identified a class of metamaterials supporting a novel mechanism of resonant electromagnetic transparency, which is different from Fano interference typically encountered in the plasmonic and metamaterial systems mimicking electromagnetically-induced transparency. It exploits destructive interference between spatially and spectrally collocated and coherently oscillating induced electric and toroidal dipoles, and produces very narrow and characteristically symmetric Lorentzian transparency lines with Q-factors exceeding 300. Such mode of the metamaterial resonant excitation corresponds to a long-awaited implementation of the non-trivial non-radiating charge-current configuration, and is expected to generate waves of gauge-irreducible vector potential in the absence of scattered electromagnetic fields.

**Acknowledgments**

This work is supported by the Royal Society and the UK's Engineering and Physical Sciences Research Council through the Nanostructured Photonic Metamaterials Programme Grant and Career Acceleration Fellowship (V.A.F.).

* e-mail: vaf@orc.soton.ac.uk